\documentclass[prb,letterpaper,aps,floatfix,twocolumn]{revtex4-1}
\usepackage{graphicx}
\usepackage{amsmath}
\usepackage{amsfonts}
\usepackage[small,bf]{subfigure}
\newcommand{\beq}{\begin{equation}}
\newcommand{\eeq}{\end{equation}}
\newcommand{\bk}{{{\bf{k}}}}

\newcommand{\bQ}{{{\bf{Q}}}}
\newcommand{\br}{{{\bf{r}}}}

\newcommand{\beqa}{\begin{eqnarray}}
\newcommand{\eeqa}{\end{eqnarray}}
\newcommand{\pdg}{{\vphantom \dag}}
\newcommand{\dg}{{\dag}}
\newcommand{\bnabla}{{\boldsymbol \nabla}} 
\newcommand{\bsigma}{{\boldsymbol \sigma}}

\newcommand{\upa}{\uparrow}
\newcommand{\da}{\downarrow} 
\newcommand{\ra}{\rightarrow}

\newcommand{\cL}{{\cal L}}
\begin{document}
\title{Charge density waves in Weyl semimetals}
\author{Dan Sehayek}
\affiliation{Department of Physics and Astronomy, University of Waterloo, Waterloo, Ontario 
N2L 3G1, Canada} 
\author{Manisha Thakurathi}
\affiliation{Department of Physics and Astronomy, University of Waterloo, Waterloo, Ontario 
N2L 3G1, Canada} 
\author{A.A. Burkov}
\affiliation{Department of Physics and Astronomy, University of Waterloo, Waterloo, Ontario 
N2L 3G1, Canada} 
\date{\today}
\begin{abstract}
We present a theory of charge density wave (CDW) states in Weyl semimetals and their interplay with the chiral anomaly. 
In particular, we demonstrate a special nature of the shortest-period CDW state, which is obtained when the separation between the Weyl nodes 
equals exactly half a primitive reciprocal lattice vector. Its topological properties are shown to be distinct from all other Weyl CDW states. 
We make a connection between this observation and the three-dimensional fractional quantum Hall state, which was recently 
proposed to exist in magnetic Weyl semimetals. 
\end{abstract}
\maketitle
\section{Introduction}
\label{sec:1}
Our understanding of topological phases of matter relies on the concept of quantum anomalies. 
Anomalies were originally discovered in the particle physics context,~\cite{Adler69,Jackiw69}
and have historically been described in terms of ``violation of classical symmetry by quantum effects". 
In the context of topological phases of condensed matter, the meaning of the term is somewhat more general: 
anomaly means a topological obstruction to a gapped symmetric nondegenerate ground state.~\cite{Volovik03,Volovik07,Wen13,Witten16}
The simplest example of this is the electron filling per unit cell of a crystal, odd versus even, which determines whether the material is 
a metal or an insulator (or an accidental semimetal). The corresponding anomaly is the chiral anomaly and its topological nature
is behind the universality of the Luttinger volume in metals.~\cite{Luttinger60,Oshikawa00,Hastings04} 
More recently discussed examples are gapless surface states of topological insulators (TI),~\cite{Haldane88,Hasan10,Qi11}
and gapless bulk states of Weyl and Dirac semimetals.~\cite{Weyl_RMP,Burkov_ARCMP,Felser_ARCMP,Hasan_ARCMP}

Of particular interest is the interplay of the anomalies with electron-electron interactions.
Even though topological nature of the anomaly gives the corresponding gapless state some degree of immunity from the effect of the interactions (or disorder), 
interactions that are strong enough may defeat the anomaly-mandated gaplessness. 
Since the logical path from the anomaly to gaplessness relies on gauge invariance and quantization of the electron quantum numbers, one way the gaplessness may be 
avoided is via fractionalization of the quantum numbers and the formation of a gapped topologically ordered state. 
In the context of Weyl semimetals this has recently been explored in Refs.~\onlinecite{Meng16,Morimoto16,Sagi18,Meng19,Teo19,Wang20,Thakurathi20}. 
In particular in Refs.~\onlinecite{Wang20,Thakurathi20} we have shown that one may open a gap in a magnetic Weyl semimetal without breaking any symmetries and 
the resulting state is a three-dimensional (3D) generalization of the familiar 2D fractional quantum Hall effect (FQHE). 

The 3D fractional quantum Hall (FQH) state in Weyl semimetals is interesting for a number of reasons. The standard 2D FQHE relies on the existence of dispersionless Landau levels. 
This dispersionless property does not easily generalize to 3D: if one, for example, makes a stack of coupled 2D FQH liquids, the Landau levels 
will inevitably acquire dispersion in the stacking direction, which will result in a metal at fractional filling factors, whose ultimate fate in the presence of interactions is not 
obvious.~\cite{Levin09}
Our proposal relies on gapping band touching points rather than fractionally filling flat Landau levels (or their generalization, bands with nontrivial Chern numbers) 
and in this sense is somewhat related to a recent proposal of a ``fractional excitonic insulator" in 2D.~\cite{Kane18}

Landau levels aside, there is a deeper reason to be skeptical that FQHE can be generalized to 3D. 
The established theoretical picture of the 2D FQHE relies on the idea of transmutability of the exchange statistics and the existence of anyons, which is a strictly 2D phenomenon,
only bosons and fermions are possible in 3D.
We have demonstrated however, that FQHE is still possible in 3D, even though quasiparticle excitations are only bosons or fermions.~\cite{Wang20,Thakurathi20}
Anyons, which are essential for FQHE, are hidden in this case, but are revealed as intersections of vortex loop excitations with crystalline topological defects,
such as the extra half atomic plane of an edge dislocation.  

Interactions may also affect gapless anomaly-mandated states in less exotic ways. 
In particular, gaplessness may be eliminated by symmetry breaking, such as the formation of a charge density wave (CDW), which breaks crystal 
translational symmetry and changes the size of the unit cell. 
Recent experiments have in fact identified (TaSe$_4$)$_2$I as a likely Weyl CDW material.~\cite{Wang_CDW,Felser_CDW}
Our present work is motivated in part by these experiments (see Refs.~\onlinecite{Ran11,Roy15,Rachel16,Ye16,Roy17,Li19,Bradlyn20} for earlier theoretical work on CDW states in Weyl semimetals). One of our goals here is to connect certain properties of the Weyl CDW states, that have been overlooked before, and the 3D FQH state, proposed 
in Refs.~\onlinecite{Wang20,Thakurathi20}. 

Weyl semimetal is a gapless topological phase protected by translational symmetry. 
A common starting point when modeling Weyl semimetals is a low-energy Hamiltonian with an even number, two in the simplest case, of independent linearly-dispersing 
species of Weyl fermions, with an equal number of right-handed (R) and left-handed (L) fermions. 
In this low-energy picture crystal translational symmetry acts as a continuous chiral symmetry, which is ``violated" by the chiral anomaly. 
The formation of a CDW state breaks this continuous chiral symmetry.  
An important manifestation of the chiral anomaly is the appearance of gapless 1D chiral modes in the vortex core of the CDW order parameter,~\cite{Zhang13,Chamon16}
which was first pointed out in a different context by Callan and Harvey.~\cite{CallanHarvey}
These chiral modes, or ``axion strings", lead to a topological term in the nonlinear sigma model (NLSM), governing the fluctuations of the phase of the CDW order parameter, 
which makes it impossible to obtain a gapped symmetric state by disordering the phase.~\cite{AbanovWiegmann,Mudry15}

This picture, however, is oversimplified, since the physical crystal translational symmetry, broken by the CDW, is not continuous. 
In this paper we show that this does in certain cases lead to significant qualitative differences from the picture based on low-energy continuum models. 
Particularly interesting turns out to be the case when the separation between the Weyl nodes is half the primitive reciprocal lattice vector, which leads to the smallest 
period, i.e. double the original lattice constant, CDW. The order parameter is then purely real, with the phase taking only two possible values, $0$ and $\pi$.
This means that topological defects are not vortices but domain walls, which qualitatively changes the nature of the defect-bound gapless states.
We show that this provides a complementary picture of the 3D FQHE, which exists in correlated Weyl semimetals at precisely this value of the Weyl node separation. 

The rest of the paper is organized as follows. 
In Section~\ref{sec:2} we discuss mean-field theory of the CDW states in a simple lattice model of a magnetic Weyl semimetal with a pair of nodes. 
We point out a qualitative difference between the period-two CDW order parameter, obtained when the Weyl nodes are separated by half the primitive reciprocal lattice vector, and 
CDW order parameters at all other values of the node separation. 
In Section~\ref{sec:3} we discuss the consequences of this qualitative distinction for the nature of gapless bound states on topological defects of the CDW order 
parameter and the corresponding topological term in the field theory, describing its phase fluctuations. 
In Section~\ref{sec:4} we make a connection to our earlier work on the 3D FQHE in Weyl semimetals and show how this state 
may be obtained by proliferating domain walls in the period-two CDW, rather than by condensing vortices in a Weyl superconductor, which was the 
picture used in our earlier work.~\cite{Wang20,Thakurathi20}
We conclude in Section~\ref{sec:5} with a brief recap of the main results. 

\section{Mean field theory of the Weyl CDW}
\label{sec:2}
We start from the simplest lattice model of a magnetic Weyl semimetal with a pair of bands touching at two Weyl nodes, located on the $z$-axis 
in momentum space at $k_z = \pm Q$~\cite{Trivedi17}
\beq
\label{eq:1}
H_0(\bk) = \sin(k_x) \sigma_x + \sin(k_y) \sigma_y + m(\bk) \sigma_z. 
\eeq
Here the Pauli matrices $\sigma_a$ act on the band index and 
\beq
\label{eq:2}
m(\bk) = \cos(k_z) - \cos(Q) - \tilde m [2 - \cos(k_x) - \cos(k_y)].
\eeq
Eq.~\eqref{eq:1} may be viewed as a 2D Dirac Hamiltonian with a $k_z$-dependent mass.~\cite{Burkov11-1} 
The mass changes sign at $k_z = \pm Q$, which are the locations of the Weyl nodes. 
Throughout this paper we will use units in which $\hbar = c = e = a = 1$, where $a$ is the lattice constant of the simple cubic lattice on which Eq.~\eqref{eq:1} is defined. 
We will also take the hopping matrix element to be unity, i.e. measure energy in units of the hopping amplitude. 

We now add electron-electron interactions, which we take to be of the simplest Hubbard type
\beq
\label{eq:3}
H_{int} = U \sum_i \psi^\dg_{i \upa} \psi^\dg_{i \da} \psi^\pdg_{i \da} \psi^\pdg_{i \upa} = - \frac{U}{2} \sum_i (\psi^\dg_i \sigma_z \psi^\pdg_i)^2 + \ldots, 
\eeq
where $\ldots$ denote a contribution that may be subsumed into the chemical potential term. 
We take the Fermi energy to be zero, which corresponds to an ideal stoichiometric Weyl semimetal. 
Decoupling the interaction term by Hubbard-Stratonovich transformation, we obtain the imaginary time action
\beqa
\label{eq:4}
S&=&\int_0^{\beta} d \tau \left\{\sum_{\bk} \psi^\dg_{\bk} [\partial_{\tau} + H_0(\bk)] \psi^\pdg_{\bk} \right. \nonumber \\
&+& \left. \sum_i\left(\Delta_i \psi^\dg_i \sigma_z \psi^\pdg_i + 
\frac{\Delta_i^2}{2 U}\right) \right\}. 
\eeqa
Here $\Delta_i$ is a fluctuating space and time-dependent CDW order parameter (in addition to CDW, $\Delta_i$ also leads to a spin density modulation, 
but we will call it CDW for brevity). We will take it to be of the form
\beq
\label{eq:5} 
\Delta_i = \Delta \cos (2 \bQ \cdot \br_i + \varphi_i), 
\eeq
where $\Delta$ is the amplitude, which we take to be constant for simplicity, $\bQ = Q \hat z$, $\br_i$ are the Bravais lattice vectors of the simple cubic lattice
and $\varphi_i$ is a fluctuating phase of the CDW order parameter, which represents sliding motion of the CDW relative to the lattice. 
In continuum, the phase fluctuations would represent a soft Goldstone mode, arising from the breaking of the continuous translational symmetry of empty space. 
In the presence of an underlying lattice the broken symmetry is no longer continuous and the phase fluctuation mode is gapped, although it is nearly gapless away 
from the values of $Q$, corresponding to highly commensurate short-wavelength CDW.
Note that a term of the type $\Delta_i \psi^\dg_i \psi^\pdg_i$ could also be present in Eq.~\eqref{eq:4}. This could result from longer-range density-density interactions and is allowed by symmetries, but it does not open a gap, at least for small values of $\Delta$, and we will not include it for this reason. 

Let us start from a mean-field theory of the CDW, which corresponds to taking $\varphi_i = \varphi$ to be a constant. 
The problem then reduces to diagonalizing the following mean-field Hamiltonian
\beq
\label{eq:6}
H = \sum_{\bk} \psi^\dg_{\bk} H_0(\bk) \psi^\pdg_{\bk} + \frac{\Delta}{2} \sum_{\bk} \left(\psi^\dg_{\bk + 2 \bQ} \sigma_z \psi^\pdg_{\bk} e^{i \varphi} + h.c. \right). 
\eeq
For $2 Q$ equal to any rational fraction of $2 \pi$ this Hamiltonian is diagonalized by folding into the reduced Brillouin zone (BZ), with $k_z$ restricted to the 
interval $-Q \leq k_z < Q$. 
Let us first explicitly solve the simplest case with $2 Q = \pi$, i.e. exactly half the original BZ size, corresponding to the shortest-period CDW with a doubled unit cell. 

In this case, Eq.~\eqref{eq:6} may be written as
\beqa
\label{eq:7}
H&=&\frac{1}{2} \sum_{\bk} \left[\psi^\dg_{\bk + 2 \bQ} H_0(\bk + 2 \bQ) \psi^\pdg_{\bk + 2 \bQ} + \psi^\dg_{\bk} H_0(\bk) \psi^\pdg_\bk \right. \nonumber \\
&+&\left. \Delta \left(\psi^\dg_{\bk + 2 \bQ} \sigma_z \psi^\pdg_\bk e^{i \varphi} + \psi^\dg_{\bk - 2 \bQ} \sigma_z \psi^\pdg_\bk e^{- i \varphi} \right) \right]. 
\eeqa
When $2 Q = \pi$, we have $\bk - 2 \bQ = \bk + 2 \bQ \,\, \textrm{mod} \,\, 2 \pi$, which gives
\beqa
\label{eq:8}
H&=&\sum_{\bk} \left[\psi^\dg_{\bk + 2 \bQ} H_0(\bk + 2 \bQ) \psi^\pdg_{\bk + 2 \bQ} +\psi^\dg_\bk H_0(\bk) \psi^\pdg_\bk \right. \nonumber \\
&+&\left. \Delta \cos(\varphi) (\psi^\dg_{\bk + 2 \bQ} \sigma_z \psi^\pdg_\bk + h.c.) \right],
\eeqa
where $\bk$ is now restricted to the reduced BZ with $-Q \leq k_z < Q$. 
Introducing a four-component spinor $\tilde \psi_{\bk} = (\psi_{\bk + 2 \bQ}, \psi_{\bk})$, the Hamiltonian may be rewritten as
\beqa
\label{eq:9} 
H&=&\sum_\bk \tilde \psi^\dg_\bk \left[\sin(k_x) \sigma_x + \sin(k_y) \sigma_y \right. \nonumber \\
&-&\left.\tilde m(2 - \cos(k_x) - \cos(k_y)) \sigma_z + \tau_z \sigma_z \cos(k_z) \right. \nonumber \\
&+& \left.\Delta \cos(\varphi) \tau_x \sigma_z \right] \tilde \psi^\pdg_{\bk},
\eeqa
where the Pauli matrices $\tau_z$ act on the two extra components of the four-spinor $\tilde \psi_{\bk}$. 
Diagonalizing one obtains the band dispersion
\beq
\label{eq:10}
\epsilon_{r s}(\bk) = s \sqrt{\sin^2(k_x) + \sin^2(k_y) + m_r^2(\bk)}, 
\eeq
where $r, s = \pm$ and 
\beqa
\label{eq:11}
m_r(\bk)&=&- \tilde m [2 - \cos(k_x) - \cos(k_y)] \nonumber \\
&+&r \sqrt{\cos^2(k_z) + \Delta^2 \cos^2(\varphi)}. 
\eeqa
Thus, even though Eq.~\eqref{eq:9} has the appearance of a 3D Dirac Hamiltonian, it is in fact not, as obvious from Eqs.~\eqref{eq:10} and \eqref{eq:11}: 
all bands are nondegenerate due to broken time-reversal (TR) symmetry. 
The band dispersion is fully gapped for all $\varphi \neq \pm \pi/2$ and the gap is maximal when $\varphi = 0, \pi$. 
Thus it is clear that $\varphi = 0, \pi$ are the energetically preferred values of the phase of the CDW order parameter. 
The CDW ground state is two-fold degenerate, with the two states related to each other by a half-translation with respect to the doubled primitive translation vector. 

It is straightforward to generalize this result to Weyl node separation $2 Q$ which is an arbitrary rational fraction of the reciprocal lattice vector $2 \pi$. 
In this case there always exists an integer $N$, such that $\bk + 2 N \bQ = \bk\,\, \textrm{mod} \,\, 2 \pi$, and the mean-field Hamiltonian may be written as
\beqa
\label{eq:12}
H&=&\sum_{\bk} \sum_{n = 0}^{N-1} \left[\psi^\dg_{\bk + 2 n \bQ} H_0(\bk + 2 n \bQ) \psi^\pdg_{\bk + 2 n \bQ} \right. \nonumber \\
&+&\left. \frac{\Delta}{2} \left(\psi^\dg_{\bk + 2 n \bQ + 2 \bQ} \sigma_z \psi^\pdg_{\bk + 2 n \bQ} e^{i \varphi} + h.c. \right)\right],
\eeqa
where $-Q \leq k_z < Q$.
Importantly since $\bk + 2 N \bQ = \bk\,\, \textrm{mod} \,\, 2 \pi$, we have $\bk + 2(N-1)\bQ + 2 \bQ = \bk \,\, \textrm{mod} \,\, 2 \pi$. 
It follows that momentum $\bk$ is coupled not only to $\bk + 2 \bQ$, but also to $\bk + 2 (N-1) \bQ$ and the momentum-space Hamiltonian 
takes the following matrix form
\begin{widetext}
\beqa
\label{eq:13}
H(\bk) = \left(
\begin{array}{ccccc}
H_0(\bk) & \frac{\Delta}{2} e^{- i \varphi} \sigma_z & 0 & \ldots & \frac{\Delta}{2} e^{i \varphi} \sigma_z \\
\frac{\Delta}{2} e^{i \varphi} \sigma_z & H_0(\bk + 2 \bQ) & \frac{\Delta}{2} e^{- i \varphi} \sigma_z & \ldots  & 0 \\
0 & \frac{\Delta}{2} e^{i \varphi} \sigma_z & H_0(\bk + 4 \bQ) & \ldots & 0 \\
\vdots & \vdots & \vdots & \ddots & \vdots \\
\frac{\Delta}{2} e^{- i \varphi} \sigma_z & \ldots & \ldots & \frac{\Delta}{2} e^{i \varphi} \sigma_z & H_0[\bk + 2 (N-1) \bQ] 
\end{array}
\right). 
\eeqa
\end{widetext}
In this case one finds that the bandgap is a function of $\Delta^N \cos(N \varphi)$ and is maximized when $\varphi = 2 \pi n/N$, $n = 0, \ldots, N -1$. 
As $N$ increases, the low-energy electronic structure near the Weyl nodes is better and better approximated by the upper $2 \times 2$ block of the 
Hamiltonian, which reduces to the standard low-energy model of left- and right-handed Weyl fermions, coupled by a complex mass term 
$\Delta \sigma_z [\cos(\varphi) \tau_x + \sin(\varphi) \tau_y]$. 
Note the important difference from the $N = 2$ case, where the mass term is real.

\section{Chiral anomaly and zero-energy bound states}
\label{sec:3}
Now let us go back to the $2 Q = \pi$ case and consider a domain wall between the two degenerate CDW states, corresponding to $\varphi = 0, \pi$. 
To find an analytical solution for the domain wall bound state it is convenient to start from the following unitary transformation of the 
momentum-space Hamiltonian in Eq.~\eqref{eq:9}
\beq
\label{eq:14}
\tau_x \ra \tau_x, \,\, \tau_y \ra - \tau_z, \,\, \tau_z \ra \tau_y, 
\eeq
followed by
\beq
\label{eq:15}
\tau_{x,y} \ra \sigma_z \tau_{x,y}, \,\, \sigma_{x,y} \ra \tau_z \sigma_{x,y}. 
\eeq
This brings the Hamiltonian to the form 
\beqa
\label{eq:16}
&&H(\bk) = \sin(k_x) \tau_z \sigma_x  + \sin(k_y)  \tau_z \sigma_y + \cos(k_z) \tau_y \nonumber \\
&+&\Delta \cos(\varphi) \tau_x - \tilde m [2 - \cos(k_x) - \cos(k_y)] \sigma_z.
\eeqa
This looks like a 3D Dirac Hamiltonian with an extra TR-symmetry breaking term (the last one). 

Now consider a domain wall, such that $\varphi(z \ra -\infty) = \pi$ and $\varphi(z \ra \infty) = 0$. 
Expanding $H(\bk)$ to linear order around the Dirac point at $k_z = \pi/2$, replacing $k_z = - i  \partial/ \partial z$, one obtains the zero-energy Jackiw-Rebbi 
soliton~\cite{JackiwRebbi} solution
\beq
\label{eq:17}
\Psi(z) = e^{ - \Delta \int_0^z d z' \cos[\varphi(z')]} | \tau_z = -1 \rangle. 
\eeq
It follows that, at a general $k_{x,y}$ the domain wall bound state is described by the following massless 2D Dirac Hamiltonian
\beqa
\label{eq:18}
H_{2D}(\bk)&=&- \sin(k_x) \sigma_x - \sin(k_y) \sigma_y \nonumber \\
&-&\tilde m [2 - \cos(k_x) - \cos(k_y)] \sigma_z. 
\eeqa
The sign of the first two terms flips if the phase changes in the opposite direction, i.e. from $0$ to $\pi$. 
Note that this does not change the Hall conductivity, associated with this 2D interface state, which is given by
$\sigma_{xy} = \textrm{sign}(\tilde m)/ 4 \pi$. 

The appearance of this 2D Dirac domain wall bound state may also be understood from the viewpoint of the anomalies. 
In the case of a noninteracting magnetic Weyl semimetal with a pair of nodes separated by a vector $2 \bQ$, chiral anomaly 
implies the following topological, thermal equilibrium contribution to the electromagnetic response~\cite{Zyuzin12-1}
\beq
\label{eq:19}
\cL_{top} = - \frac{i}{4 \pi^2} \epsilon_{\mu \nu \lambda \rho} Q_{\mu} A_{\nu} \partial_{\lambda} A_{\rho},
\eeq
where $\cL_{top}$ is the imaginary time Lagrangian density. 
When translational symmetry is spontaneously broken and a CDW gap is opened, it is usually assumed that this changes to~\cite{Zhang13}
\beq
\label{eq:20}
\cL_{top} = - \frac{i}{8 \pi^2} \epsilon_{\mu \nu \lambda \rho} (2 Q_{\mu} + \partial_{\mu} \varphi)  A_{\nu} \partial_{\lambda} A_{\rho}, 
\eeq
where $\varphi$ is the phase of the CDW order parameter, introduced above. 
This result is most easily obtained from a low-energy model of a Weyl semimetal with a pair of nodes
\beqa
\label{eq:21}
S&=&\int_0^{\beta} d \tau \int d^3 r \left[\psi^\dg_R (\partial_{\tau} - i \bnabla \cdot \bsigma) \psi^\pdg_R \right. \nonumber \\
&+&\left.\psi^\dg_L (\partial_{\tau} + i \bnabla \cdot \bsigma) \psi^\pdg_L + \frac{\Delta}{2}\left(\psi^\dg_R \psi^\pdg_L e^{i \varphi} + h.c. \right) \right]. \nonumber \\
\eeqa
After a gauge transformation $\psi_R \ra \psi_R e^{i \varphi/2}$ and $\psi_L \ra \psi_L e^{- i \varphi/2}$, this becomes
\begin{widetext}
\beqa
\label{eq:22}
S&=&\int_0^{\beta} d \tau \int d^3 r \left[\psi^\dg_R \left(\partial_{\tau} + \frac{i}{2} \partial_{\tau} \varphi - i \bnabla \cdot \bsigma + \frac{1}{2} \bnabla \varphi \cdot \bsigma\right) 
\psi^\pdg_R  + \psi^\dg_L \left(\partial_{\tau} - \frac{i}{2} \partial_{\tau} \varphi + i \bnabla \cdot \bsigma + \frac{1}{2} \bnabla \varphi \cdot \bsigma \right)  \psi^\pdg_L \right. \nonumber \\
&+&\left. \frac{\Delta}{2}\left(\psi^\dg_R \psi^\pdg_L + \psi^\dg_L \psi^\pdg_R \right) \right],
\eeqa
\end{widetext}
from which Eq.~\eqref{eq:20} follows since the first two terms in \eqref{eq:22} describe a Weyl semimetal with a pair of nodes, separated by the vector $\bnabla \varphi$ in momentum space. 

This logic is correct, except when $2 Q = \pi$. In this case the mass term, coupling the left- and right-handed Weyl fermions, is real, unlike in Eq.~\eqref{eq:21}. 
Its phase
\beq
\label{eq:23}
\theta = \frac{\pi}{2} \left[1 - \textrm{sign}(\cos(\varphi))\right],
\eeq
can thus only take two values, $0$ and $\pi$. Its contribution to the Lagrangian then takes the form, which appears identical 
to a 3D TR-invariant TI
\beq
\label{eq:24}
\cL_{top} = - \frac{i \theta}{8 \pi^2} \epsilon_{\mu \nu \lambda \rho} \partial_{\mu} A_{\nu} \partial_{\lambda} A_{\rho}. 
\eeq
This follows from the fact that the Dirac Hamiltonian Eq.~\eqref{eq:16}, when expanded to linear order around the gapped Dirac point at $k_x = k_y = 0, k_z = \pi/2$, is identical 
to the low-energy Hamiltonian of a 3D TI. 
Eq.~\eqref{eq:24} may then be obtained by standard arguments, for example using the Fujikawa's method,~\cite{Fujikawa79} while applying a sequence of infinitesimal chiral transformations to the linearized Dirac Hamiltonian to change the sign of the mass term,~\cite{Hosur10,Zyuzin12-1} thus transforming between an ordinary insulator and a 3D TI. 
This similarity to the 3D TI makes it tempting to identify a gapped Weyl semimetal at $2 Q = \pi$ and $\theta = \pi$ with an {\em axion insulator},~\cite{Wan11,Zhang13,You16}
in which TR is broken but there is a quantized magnetoelectric response due to the still well-defined and quantized $\theta$. 
However, such an identification would not really be correct. 
In the case of a true axion insulator, $\theta = \pi$ and $\theta = 0$ correspond to two topologically-distinct states, i.e. an axion insulator and an ordinary TR-broken 
insulator. They are distinguished by a quantized magnetoelectric response,~\cite{Wan11,Essin09} as well as presence or absence of chiral hinge states.~\cite{Varnava18} 
In contrast, the $\theta = 0, \pi$ states of a magnetic Weyl semimetal, gapped by a period-two CDW, are related to each other by a crystal symmetry operation, 
i.e. a half-CDW-period translation, and already for this reason can not be topologically distinct. This observation is in agreement with Ref.~\onlinecite{Bradlyn20}, 
which has also recently explored manifestations of the lattice-scale physics in the CDW states in Weyl semimetals. 

What about the massless 2D Dirac bound state that one obtains at a domain wall between the $\theta = 0$ and $\theta = \pi$ CDW insulators? 
Recall that a 2D lattice-regularized Dirac fermion of Eq.~\eqref{eq:18} corresponds to a critical point between 2D insulators with $\sigma_{xy} = 0$ and 
$\sigma_{xy} = 1/2 \pi$ and thus produces a half-quantized Hall conductivity $\sigma_{xy} = \textrm{sign}(\tilde m) /4\pi = 1/4 \pi$.~\cite{Ludwig94}
Such a half-quantized Hall conductivity {\em per atomic plane} is identical to the Hall conductivity of a Weyl semimetal with $2 Q = \pi$, 
$\sigma_{xy} = 2 Q/ 4 \pi^2 = 1/4 \pi$, which is preserved when the CDW gap is opened. 
This implies that a Weyl semimetal with $2 Q = \pi$ may be viewed as a stack of 2D atomic layers with $\sigma_{xy} = 1/4 \pi$, coupled in such a way that the 
Hall conductivity per layer is preserved. Indeed, as discussed above, Weyl semimetal Hamiltonian Eq.~\eqref{eq:1} has the form of a 2D Dirac Hamiltonian 
with a $k_z$-dependent mass. Even though the mass is nonzero everywhere except at the locations of the Weyl points, the contribution of low-energy states near 
the $ k_x = k_y = 0$ axis to the total 3D Hall 
conductivity is zero, since the contribution of the interval $-Q \leq k_z < Q$ is exactly cancelled by the interval $|k_z| > Q$. 
This means that only high-energy states contribute to the Hall conductivity, giving $\sigma_{xy} = \textrm{sign}(\tilde m) /4\pi = 1/4 \pi$ per each value of $k_z$. 
Since the CDW states, corresponding to $\theta = 0, \pi$ are related by a half-period translation, a domain wall between them leaves one ``unpaired" atomic plane, carrying 
$\sigma_{xy} = 1/4 \pi$ and thus a massless 2D Dirac fermion. 

The existence of this massless Dirac fermion state also follows from Eq.~\eqref{eq:24}, which, when evaluated in a sample with a domain wall between the two CDW 
states, corresponding to $\theta= 0, \pi$, gives
\beq
\label{eq:25}
\cL_{top} = - \frac{i }{8 \pi} \epsilon_{z \mu \nu \lambda} A_{\mu} \partial_{\nu} A_{\lambda}, 
\eeq
which is precisely the response of a massless 2D Dirac fermion, corresponding to $\sigma_{xy} = 1/ 4\pi$. 
Note that the sign of $\sigma_{xy}$ is undefined in this formulation due to the $2 \pi$ ambiguity of the definition of $\theta$.  

It is instructive to contrast these states with the superficially-similar surface states of a 3D TR-invariant TI. 
In this case the momentum-space Hamiltonian Eq.~\eqref{eq:16} is replaced by
\beq
\label{eq:26}
H(\bk) = \sin(k_x) \tau_z \sigma_x  + \sin(k_y)  \tau_z \sigma_y - k_z \tau_y + m(\bk) \tau_x, 
\eeq
where 
\beq
\label{eq:27}
m(\bk) = \Delta - \tilde m [2 - \cos(k_x) - \cos(k_y)]. 
\eeq
Taking the gap parameter $\Delta$ to be a function of $z$, such that $\Delta(z \ra -\infty) < 0$ and $\Delta(z \ra \infty) >  0$, one obtains, at $k_{x,y} = 0$ a zero-energy 
bound state solution that is identical to Eq.~\eqref{eq:17}
\beq
\label{eq:28}
\Psi(z) = e^{ - \int_0^z  d z' m(0,0,z') } | \tau_z = -1 \rangle. 
\eeq
Unlike Eq.~\eqref{eq:17}, however, this solution continues to exist only as long as the Dirac mass $m(k_x, k_y, z)$ actually changes sign as a function of $z$, 
which only happens at small enough $k_{x,y}$. 
The surface state then exists only in the vicinity of $k_x = k_y =0$ and is not associated with a nonzero Hall conductivity, as long as it is gapless. 
\section{Gapped strongly correlated Weyl semimetal from disordered CDW}
 \label{sec:4}
 CDW states, described above, provide an interesting complementary prospective on the question of opening a gap in a Weyl semimetal without explicitly breaking the protecting 
 symmetries. 
 We explored this question before using the ``vortex condensation" method,~\cite{Wang20,Thakurathi20} where one starts from a gapped superconducting 
 state, induced in a Weyl semimetal, and asks if it is possible to destroy the superconducting coherence while keeping the gap. 
 If successful, this procedure produces an insulator, which nevertheless preserves topological response of the underlying Weyl semimetal. 
 
Similar approach may be applied to the gapped CDW states. In this case one imagines keeping the CDW gap $\Delta$ intact, while disordering the phase $\varphi$
and thus restoring the broken translational symmetry. 
As in the case of the phase-disordered superconductor, topological defects play a crucial role here. 
In particular, as discussed in Section~\ref{sec:3}, when $2 Q \neq \pi$, the NLSM, which describes phase fluctuations of the CDW order parameter
\beq
\label{eq:29}
\cL = \frac{1}{2 g}(\partial_{\mu} \varphi)^2 + \cL_{top}, 
\eeq
where $1/g \sim \Delta^2 \ln(\Lambda/\Delta)$ and $\Lambda \gg \Delta$ is of the order of the total bandwidth, contains a topological term
\beq
\label{eq:30}
\cL_{top} = - \frac{i}{8 \pi^2} \epsilon_{\mu \nu \lambda \rho} \partial_{\mu} \varphi A_{\nu} \partial_{\lambda} A_{\rho}.
\eeq
As first shown by Callan and Harvey,~\cite{CallanHarvey} this term necessarily leads to the appearance of 1D chiral modes in the core of the vortex loops of the 
phase $\varphi$. The chirality and number of the 1D modes reflects the vorticity and the modes cross zero energy at the values of momenta, corresponding to the 
locations of the Weyl nodes,~\cite{Zhang13,Chamon16,Wang20} as may be seen by an explicit solution of the corresponding Dirac equation in the presence of a vortex. 
These 1D chiral modes in the vortex cores necessarily lead to a gapless state once the translational symmetry is restored. This is because the only way to eliminate the gapless 
chiral modes is to 
hybridize them in pairs of opposite chirality, which is impossible without breaking translational symmetry since they exist at different momenta ($\pm Q \hat z$). 
Note that the phase anisotropy, that arises due to lattice commensuration effects, as discussed in Section~\ref{sec:2}, does not change this picture. 

This is correct at all values of the Weyl node separation, except when $2 Q = \pi$. As discussed above, in this case the mass term, induced by the CDW order 
parameter, is purely real and, as a consequence, topological defects are 2D domain walls instead of 1D vortex loops. 
The corresponding topological term is given by Eq.~\eqref{eq:24}, which is very different from Eq.~\eqref{eq:30}. 
In the language of the anomalies, Eq.~\eqref{eq:30} expresses the perturbative [in the sense that $|\partial_{\mu} \varphi|$ may be arbitrarily small and thus Eq.~\eqref{eq:30} may be obtained from a perturbative gradient expansion of the imaginary time action] chiral anomaly of gapless Weyl fermions. The anomaly of Eq.~\eqref{eq:24} is instead nonperturbative, or global, and is closely related to the 2D parity anomaly [Eq.~\eqref{eq:25}, which follows 
from Eq.~\eqref{eq:24}, is in fact a direct manifestation of the 2D parity anomaly].~\cite{Witten16}
Since each domain wall binds a 2D massless Dirac fermion, the question of gapping the Weyl semimetal without breaking translational symmetry reduces in this case 
to the question of gapping a 2D Dirac fermion, while preserving its half-integer Hall conductivity $\sigma_{xy} = 1/4 \pi$. 
While a closely-related question of gapping the 2D Dirac surface states of the 3D TR-invariant TI has been discussed before,~\cite{FidkowskiSurface,Wang13,Metlitski15,Fidkowski14,Bonderson13,WangSenthil14,MetlitskiVortex,PotterWangMetlitskiVishwanath} 
we will nevertheless go through the procedure in detail. 
The procedure we use here has not been described in the literature explicitly, although it is implicit in, for example, the approach of Ref.~\onlinecite{Wang13}.  
This will also facilitate the connection to our own earlier work.~\cite{Wang20,Thakurathi20} 

Let us start from the 2D Dirac Hamiltonian of Eq.~\eqref{eq:18}, which describes the gapless bound state on a CDW domain wall at $2 Q = \pi$. 
In real space this becomes
\beqa
\label{eq:31}
H&=&\sum_\br\left[\frac{i}{2} \psi^\dg_\br (\sigma_i - i \tilde m \sigma_z) \psi^\pdg_{\br + i} e^{i A_{\br i}} + h.c. \right. \nonumber \\
&-&\left.2 \tilde m \psi^\dg_\br \sigma_z \psi^\pdg_\br + i A_{\br 0} \psi^\dg_\br \psi^\pdg_\br \right], 
\eeqa
where we have coupled the fermions to an external probe electromagnetic field $A_{\mu}$. 
We then use parton decomposition~\cite{Georges_SR}
\beq
\label{eq:32}
\psi_\br = e^{i \theta_\br} f_\br, 
\eeq
where $e^{i \theta_\br}$ annihilates a spinless boson (chargon), carrying the charge of the electron, while $f_\br$ is a neutral spinon, carrying the spin. 
The phase $\theta_\br$ is conjugate to the chargon number $n_\br$, which satisfies the constraint $f^\dg_\br f^\pdg_\br = n_\br$. 
The imaginary-time Lagrangian density (the action is $S = \int d \tau \sum_\br \cL$) then takes the following form~\cite{Lee_SR,Senthil_SR,Barkeshli12,Burkov19}
\beqa
\label{eq:33}
\cL_f&=&f^\dg_\br (\partial_{\tau} - i a_{\br 0} ) f^\pdg_\br - 2 \tilde m f^\dg_\br \sigma_z f^\pdg_\br \nonumber \\
&+&\frac{i \chi}{2} f^\dg_\br (\sigma_i - i \tilde m \sigma_z) f^\pdg_{\br + i} e^{- i a_{\br i}} + h.c.,
\eeqa
and 
\beq
\label{eq:34}
\cL_b = i n_\br (\partial_{\tau} \theta_\br + A_{\br 0} + a_{\br 0}) - \chi \cos(\Delta_i \theta_\br + A_{\br i} + a_{\br i}). 
\eeq
Here the total Lagrangian $\cL = \cL_f + \cL_b$ and $a_{\br \mu}$ is a statistical gauge field, which couples chargons and spinons. 
Eqs.~\eqref{eq:33}, \eqref{eq:34} are obtained by a Hubbard-Stratonovich decoupling of the chargons and spinons in the original electron imaginary time 
action with $a_{\br \mu}$ emerging as the phase of the Hubbard-Stratonovich field, while an approximately constant $\chi$ is its magnitude. 

Transforming the cosine by the Villain transformation, we obtain
\beq
\label{eq:35}
\cL_b = i J_{\br \mu} (\Delta_{\mu} \theta_\br + A_{\br \mu} + a_{\br \mu}) + \frac{1}{2 \chi} J^2_{\br \mu}, 
\eeq
where $\mu = 0,x,y$, $J_{\br 0} \equiv n_{\br}$ and we have included a term $n_\br^2/2 \chi$, arising from the electron-electron interactions, which have 
been implicit up to this point. The coefficient of the interaction term was taken to be $1/2 \chi$ for brevity, its specific value does not matter. 
The new variables $J_{\br \mu}$ are integers, defined on the links $(\br \mu)$ of the lattice and represent chargon space-time currents.  
Integrating out the phases $\theta_\br$, one obtains the conservation law for the chargon currents
\beq
\label{eq:36}
\Delta_\mu J_{\br \mu} = 0, 
\eeq
which may be solved as
\beq
\label{eq:37}
J_{\mu} = \frac{1}{2 \pi} \epsilon_{\mu \nu \lambda} \Delta_\nu b_\lambda, 
\eeq
where we will drop the $\br$ indices henceforth and $b_{\mu}$ are $2 \pi \mathbb{Z}$ valued variables, defined on the links of the dual lattice. 
It is convenient to soften the $2 \pi \mathbb{Z}$ constraint by introducing a vortex kinetic energy term, which brings the Lagrangian to the form
\beqa
\label{eq:38}
\cL_b&=&\frac{i}{2 \pi} (A_{\mu} + a_{\mu}) \epsilon_{\mu \nu \lambda} \Delta_{\nu} b_\lambda + \frac{1}{8 \pi^2 \chi} (\epsilon_{\mu \nu \lambda} b_{\lambda})^2 \nonumber \\
&-& t \cos(\Delta_{\mu} \phi + b_{\mu}), 
\eeqa
where $e^{i \phi}$ is an annihilation operator for a vortex carrying flux $2 \pi$ (i.e. $h c /e$ in ordinary units). 

Let us now shift our attention to the spinons. 
We assume that the spinons are paired by the usual BCS singlet $s$-wave pairing term, which opens a gap. 
Ignoring the statistical gauge field for a moment, this is described by the following momentum-space Hamiltonian
\beqa
\label{eq:39}
H&=&-\sum_\bk f^\dg_\bk [\chi \sigma_x \sin(k_x) + \chi \sigma_y \sin(k_y) + \sigma_z m(\bk)] f^\pdg_\bk \nonumber \\
&-& \Delta \sum_\bk(f^\dg_{\bk \upa} f^\dg_{- \bk \da} + f^\pdg_{- \bk \da} f^\pdg_{\bk \upa}), 
\eeqa
where $m(\bk) = \tilde m [2 - \cos(k_x) - \cos(k_y)]$. 
Introducing a Nambu spinor $\tilde f_\bk = (f^\pdg_{\bk \upa}, f^\pdg_{\bk \da}, f^\dg_{-\bk \da}, f^\dg_{- \bk \upa})$, the Hamiltonian reduces to a block-diagonal from
\beqa
\label{eq:40}
H&=&-\frac{1}{2} \sum_\bk \tilde f^\dg_\bk \left\{\chi \sigma_x \sin(k_x) + \chi \sigma_y \sin(k_y) \right. \nonumber \\
&+& \left. [m(\bk) \pm \Delta] \sigma_z\right\}\tilde f^\pdg_\bk. 
\eeqa
This describes a topological $p + ip$ superconductor with a chiral Majorana edge mode and a zero-energy Majorana bound state in the $h c /2 e = \pi$-flux vortex core. 
Coupling the paired spinons to the statistical gauge field $a_\mu$ produces a Meissner term for $a_\mu$, which has the form $-\cos(2 a_\mu)$ since a spinon pair 
carries charge 2 of the statistical gauge field. This makes $a_{\mu}$ a $\mathbb{Z}_2$ gauge field. Its nontrivial excitations (visons) carry flux $\pi$, which implies that a 
single vison always induces a Majorana zero-energy bound state. 

Now let us return to the dualized chargon Lagrangian Eq.~\eqref{eq:38}, and analyze possible gapped insulator phases of our system, which may be 
obtained within this formalism. 
The simplest one is obtained when we condense flux $2 \pi$ vortices, annihilated by $e^{i \phi}$. 
This produces a Higgs mass term for the gauge field $b_{\mu}$, which gaps all charged excitations. 
The spinons are also gapped by pairing, but the visons may in principle be either gapped or condensed. 
If it was possible to condense visons, this would result in an ordinary band insulator, since the fluctuating $\pi$-flux would 
bind the spinons and chargons into electrons. Vison condensation is impossible, however, since a $\pi$-flux vortex has a zero-energy Majorana 
bound state in its core, as discussed above. This is a manifestation of the nontrivial topology of the massless Dirac fermion (parity anomaly), 
which survives in the strongly-correlated state as parity anomaly of the spinon band structure. 
The state with a gapped vison has $\mathbb{Z}_2$ topological order, and is a Kitaev spin liquid.~\cite{Kitaev06}
It has a half-quantized thermal Hall conductivity 
\beq
\label{eq:41}
\kappa_{xy} = \frac{L T}{4 \pi}, 
\eeq
where $L = \pi^2 k_B^2/3$ is the Lorenz number, but zero electrical Hall conductivity. 

To obtain an insulator with the same topological response as a massless Dirac fermion, we need a state with a half-quantized thermal and electrical Hall conductivity. 
This can not be obtained by putting chargons in the $\nu = 1/2$ FQH liquid (the resulting state is the Moore-Read Pfaffian~\cite{Read00}) since the thermal Hall conductivity of this state is $\kappa_{xy} = 3 L T/ 4 \pi$, the extra quantum coming from the chiral boson edge mode of the $\nu = 1/2$ Laughlin liquid. 
The correct state is obtained instead by assuming {\em double}, i.e. flux $4 \pi$, {\em vortices}, form the $\nu = 1/2$ liquid. As can be seen by a direct inspection of the equations below, putting flux 
$2 \pi$ vortices in any quantum Hall state may only produce a state with an integer Hall conductivity. 
In Ref.~\onlinecite{Wang13} the incompatibility of the flux $2 \pi$ vortices with the half-quantized electrical Hall conductivity was instead related to the fact that such vortices have semionic exchange statistics when the Hall conductivity is half-integer. The two viewpoints are of course equivalent. 

To describe the state with $4 \pi$ vortices forming the $\nu = 1/2$ Laughlin liquid we first replace the single-vortex kinetic energy term by a double-vortex one $- t \cos(2 \Delta_\mu \phi + 2 b_\mu)$ and then apply the Villain transform
\beqa
\label{eq:42}
\cL_b&=&\frac{i}{2 \pi} (A_{\mu} + a_{\mu}) \epsilon_{\mu \nu \lambda} \Delta_{\nu} b_\lambda + \frac{1}{8 \pi^2 \chi} (\epsilon_{\mu \nu \lambda} b_{\lambda})^2 \nonumber \\
&+&2 i \tilde J_\mu (\Delta_{\mu} \phi + b_\mu) + \frac{1}{2 t} \tilde J^2_\mu,
\eeqa
where $\tilde J_\mu$ are integer-valued vortex currents. 
Integrating out $\phi$, we obtain the vorticity conservation law
\beq
\label{eq:43}
\Delta_\mu \tilde J_\mu = 0, 
\eeq
which may be solved as
\beq
\label{eq:44}
\tilde J_\mu = \frac{1}{2 \pi} \epsilon_{\mu \nu \lambda} \Delta_\nu \tilde b_\lambda, 
\eeq
where $\tilde b_\mu$ is a $2 \pi \mathbb{Z}$-valued gauge field. 
Note that $2 \pi$ flux of $\tilde b_\mu$ corresponds to a $4 \pi$ vortex. 
Then we place the double vortices in the $\nu = 1/2$ Laughlin state. Taking the continuum limit, this leads to the following Lagrangian
\beq
\label{eq:45}
\cL_b = \frac{i}{2 \pi} (A_\mu + a_\mu + 2 \tilde b_\mu) \epsilon_{\mu \nu \lambda} \partial_\nu b_\lambda - \frac{2 i }{4 \pi} \epsilon_{\mu \nu \lambda} \tilde b_\mu \partial_\nu 
\tilde b_\lambda,
\eeq
where we have ignored Maxwell terms for $b_\mu$ and $\tilde b_\mu$, which are not essential here. 

To understand the physics of the state we have obtained, let us ignore the coupling to spinons for a moment and make a variable change
$b_\mu \ra (b_\mu + \tilde b_\mu)/2$. 
Then we obtain
\beq
\label{eq:46}
\cL_b = \frac{i}{4 \pi} \epsilon_{\mu \nu \lambda} (b_\mu \partial_\nu \tilde b_\lambda + \tilde b_\mu \partial_\nu b_\lambda)
+ \frac{i}{4 \pi} A_\mu \epsilon_{\mu \nu \lambda} \partial_{\nu} (b_\lambda + \tilde b_\lambda). 
\eeq
This describes an integer quantum Hall state of two-component charge-$1/2$ bosons.~\cite{Lu-Vishwanath,Senthil-Levin}
Making another variable change
\beq
\label{eq:47}
b_\mu = c_\mu + \tilde c_\mu, \,\, \tilde b_\mu = c_\mu - \tilde c_\mu, 
\eeq
we obtain
\beq
\label{eq:48}
\cL_b = \frac{2 i}{4 \pi} \epsilon_{\mu \nu \lambda} (c_\mu \partial_\nu c_\lambda - \tilde c_\mu \partial_\nu \tilde c_\lambda) + 
\frac{i}{2 \pi} A_\mu \epsilon_{\mu \nu \lambda} \partial_\nu c_\lambda. 
\eeq
By a standard argument,~\cite{Wen_QFT} this leads to a pair of opposite-chirality edge modes: one charged, which upon integrating out the gauge field $c_\mu$ gives the half-quantized Hall conductivity 
$\sigma_{xy} = 1/4 \pi$, and one neutral, which cancels the contribution of the charged mode to thermal Hall conductivity. This state thus has a half-quantized electrical 
and zero thermal Hall conductivity. 

Now let us go back to Eq.~\eqref{eq:45} and add the spinon contribution. The total Lagrangian is given by
\beqa
\label{eq:49}
\cL&=&\cL_f(-a_\mu) + \frac{i}{2 \pi} (A_\mu + a_\mu + 2 \tilde b_\mu) \epsilon_{\mu \nu \lambda} \partial_\nu b_\lambda \nonumber \\
&-&\frac{2 i}{4 \pi} \epsilon_{\mu \nu \lambda} \tilde b_\mu \partial_\nu \tilde b_\lambda. 
\eeqa
Intregrating out $b_\mu$ one obtains at low energies
\beq
\label{eq:50}
\tilde b_\mu = - \frac{A_\mu + a_\mu}{2}. 
\eeq
Since $a_\mu$ is made a $\mathbb{Z}_2$ gauge field by spinon pairing, Eq.~\eqref{eq:50} tells us that $\tilde b_\mu$ is a $\mathbb{Z}_4$ gauge field, 
which corresponds to fractionalization of the electron as 
\beq
\label{eq:51}
\psi = b_1 b_2 f, 
\eeq
where $b_{1,2}$ are the charge-$1/2$ bosons of Eq.~\eqref{eq:46} and $f$ is the neutral spinon. 
Plugging this back into the Lagrangian, we obtain
\beqa
\label{eq:52}
\cL&=&\cL_f(-a_\mu) - \frac{i}{8 \pi} \epsilon_{\mu \nu \lambda} A_\mu \partial_\nu A_\lambda - \frac{i}{4 \pi} \epsilon_{\mu \nu \lambda} A_\mu \partial_\nu a_\lambda \nonumber \\
&-&\frac{i}{8 \pi} \epsilon_{\mu \nu \lambda} a_\mu \partial_\nu a_\lambda. 
\eeqa
This tells us that a vison, in addition to carrying a Majorana zero mode, has charge-$1/4$. 
A $2 \pi$ vortex, which is also a gapped excitation, carries a charge-$1/2$ and is a semion. 
The state we have obtained is thus a nonabelian FQH state, which has $\sigma_{xy} = 1/4 \pi$ and $\kappa_{xy} = L T/ 4 \pi$, i.e. 
an identical topological response (parity anomaly) to a massless free Dirac fermion. 
\begin{figure}[t]
\includegraphics[width=9cm]{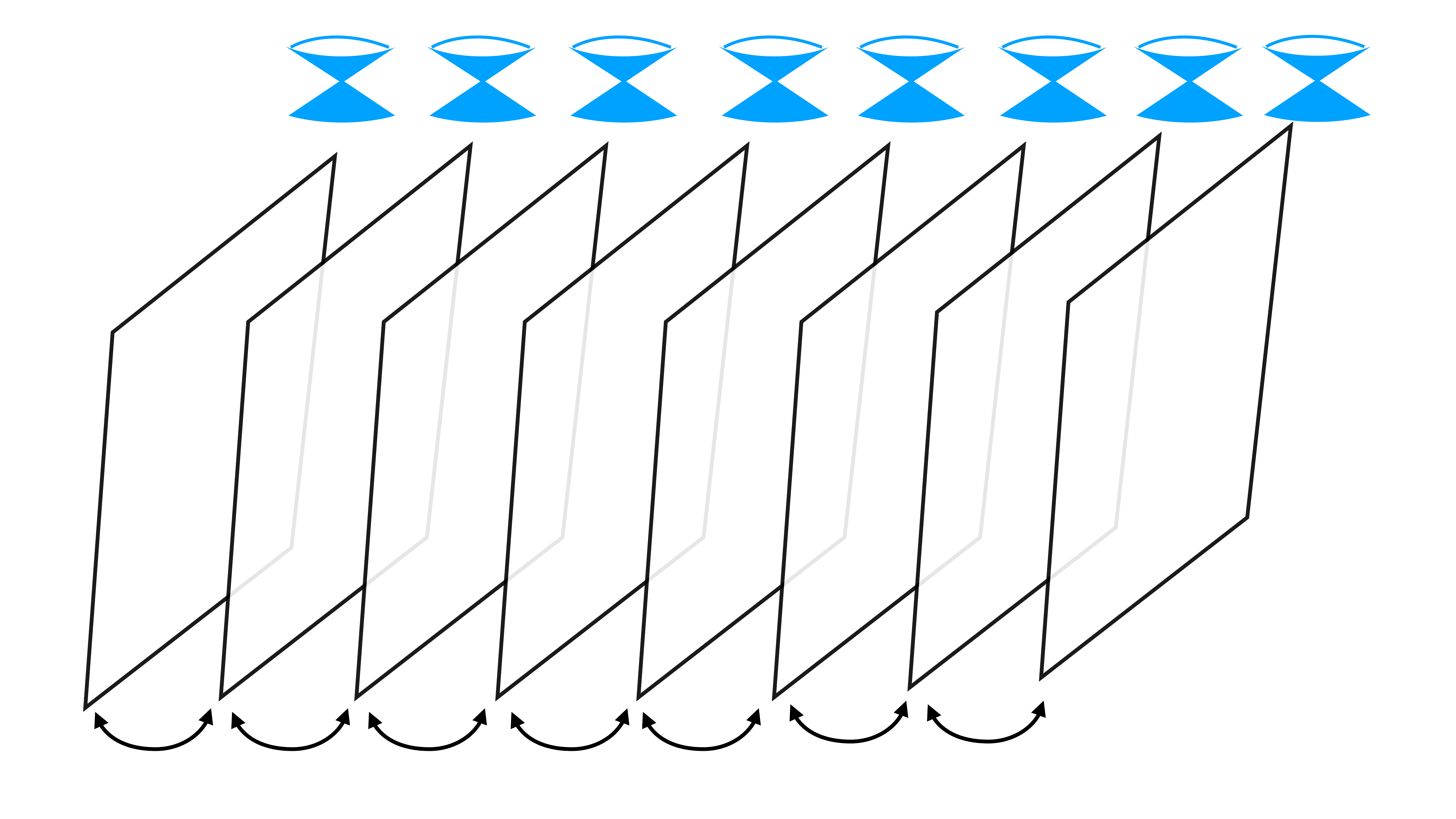}
\includegraphics[width=8cm]{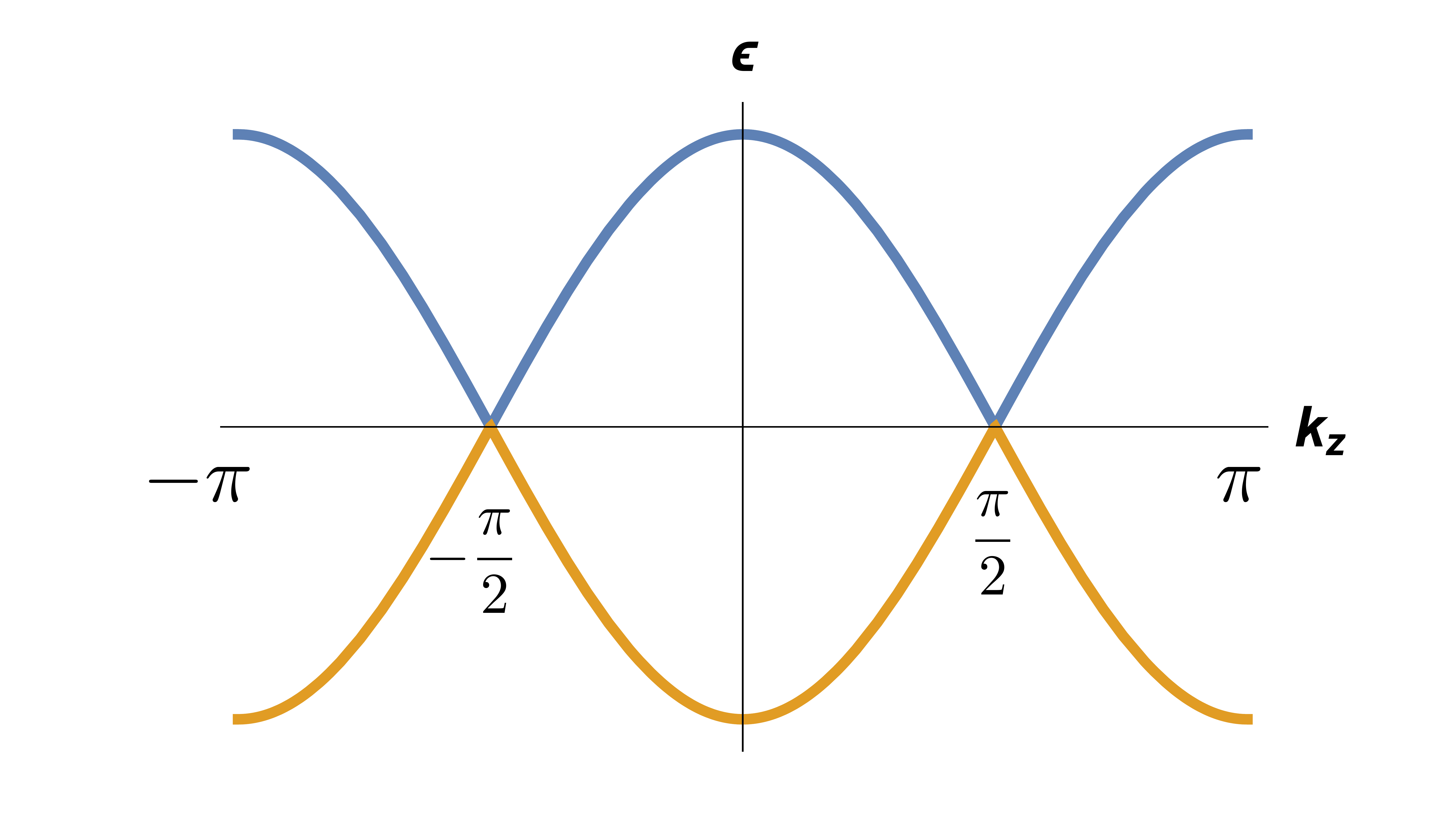}
\caption{(Color online) Weyl semimetal with $2 Q = \pi$ obtained by coupling massless 2D Dirac domain wall states.}
\label{fig:1}
\end{figure}

Stacking domain wall 2D Dirac fermion states of Eq.~\eqref{eq:18}, and allowing electron tunneling between them, clearly results in a gapless Weyl semimetal state, 
described by Eq.~\eqref{eq:16} with $\Delta = 0$, see Fig.~\ref{fig:1}. This expresses the fact that it is impossible to disorder the $2 Q = \pi$ CDW by proliferating domain walls and obtain a gapped unfractionalized insulator as a result. 
This may be viewed as a consequence of the anomaly, expressed by Eq.~\eqref{eq:24}. 
However, it is possible to obtain a gapped insulator with topological order, by stacking gapped 2D Dirac states with the $\mathbb{Z}_4$ topological order, described above. 
As a result, one obtains a 3D topologically ordered state, which may be viewed as a 3D FQH liquid, and which was described in detail in Refs.~\onlinecite{Wang20,Thakurathi20}.
The Chern-Simons theory, described above, is generalized to 3D by promoting the gauge field $b_\mu$ to a two-form antisymmetric gauge field $b_{\mu \nu}$. 
This expresses the physical fact that vortex excitations, which couple to $b_\mu$ and which are particles in 2D, become vortex loops in 3D. 
For further details we refer the reader to Refs.~\onlinecite{Wang20,Thakurathi20}. 
\section{Conclusions}
\label{sec:5}
In this paper we have discussed some aspects of the physics of CDW states in magnetic Weyl semimetals, focusing on the quantum anomalies. 
Our main result is the qualitative difference that exists between the period-two CDW, which arises when Weyl nodes, separated by half the primitive reciprocal lattice vector,
are gapped, and all other CDW states. Due to the CDW order parameter being purely real in this case, topological defects are domain walls, separating states with opposite 
sign of the order parameter. In contrast, at all other values of the Weyl node separation, the CDW order parameter is complex and topological defects are vortex loops 
(``axion strings"). This distinction has important implications for strong correlation phenomena in Weyl semimetals, in particular the question of gapping out Weyl 
nodes without explicitly breaking translational symmetry. 
We demonstrated before, using the ``vortex condensation" approach, that it is indeed possible to gap out Weyl nodes, separated by half the primitive reciprocal lattice vector, without breaking 
the translational symmetry, protecting the gapless nodes.~\cite{Wang20,Thakurathi20} The resulting state turns out to be a 3D generalization of the FQHE. 
Here we have shown how to describe the same state from a different viewpoint, that of a disordered CDW. Domain walls of the period-two CDW carry massless Dirac 
fermion bound states, which may be gapped without altering their half-quantized Hall conductivity. 
The state one obtains is a nonabelian even-denominator FQH liquid, namely the TR-broken version of the Pfaffian-antisemion state, discussed before in the context of gapped surface states of 3D TI.~\cite{FidkowskiSurface,Wang13,Metlitski15,Fidkowski14,Bonderson13,WangSenthil14,MetlitskiVortex,PotterWangMetlitskiVishwanath} 
Stacking such 2D Pfaffian-antisemion liquids corresponds to a mean-field description of the 3D FQH liquid state of Refs.~\onlinecite{Wang20,Thakurathi20}. 

In contrast, at all other values of the Weyl node separation, the topological defects of the CDW order parameter are vortices, which carry 1D chiral modes in their core. 
The chirality of the mode is determined by the sign of the vorticity and the modes cross zero energy at the momentum of the corresponding (right- or left-handed) Weyl node. 
Such chiral modes can not be gapped, except by hybridizing modes of opposite chirality, which necessarily breaks translational symmetry since modes of opposite chirality 
exist at different momenta. Thus Weyl semimetals at a general value of the Weyl node separation, not equal to half the primitive reciprocal lattice vector, may not be gapped without explicitly breaking the crystal translational symmetry. 

It is interesting to note that the special nature of the $2 Q = \pi$ Weyl semimetal is somewhat analogous to that of the half-filled 
interacting electron liquid in 1D. 
In this case the presence of Umklapp terms at half filling leads, with strong enough interactions, to an instability of the gapless Luttinger liquid and the formation of a commensurate period-two CDW. 
This analogy is not surprising, given that a Weyl semimetal with $2Q = \pi$, placed in an external magnetic field, maps via the formation of the lowest Landau level,
connecting the nodes, precisely onto a 1D metal at half filling.~\cite{Nagaosa17}

The description of the 3D FQH liquid in terms of a disordered CDW, proposed in this paper, makes it appear somewhat less exotic and more accessible, compared to the description based on a phase-incoherent superconductor. While this is of course mostly an illusion, since the two descriptions are equivalent, the recent experimental evidence for Weyl CDW in (TaSe$_4$)$_2$I~\cite{Felser_CDW} gives one some hope that the 3D FQHE may be realized experimentally in the future. A key advance needed here is a magnetic Weyl semimetal material with a pair of Weyl nodes, in which the node separation is tunable by changing the magnetization. 
\begin{acknowledgments}
We acknowledge useful discussions with C. Wang. 
DS and MT were supported by the Natural Sciences and Engineering Research Council (NSERC) of Canada.
AAB was supported by Center for Advancement of Topological Semimetals, an Energy Frontier Research Center funded by the U.S. Department of Energy Office of Science, Office of Basic Energy Sciences, through the Ames Laboratory under contract DE-AC02-07CH11358. 
\end{acknowledgments}
\bibliography{references}
\end{document}